\documentclass[prl,twocolumn,amsmath,amssymb,floatfix,superscriptaddress]{revtex4-1}
\usepackage[dvips]{graphicx}
\usepackage{amsfonts}
\usepackage{dcolumn}
\usepackage{bm}
\usepackage{color}
\usepackage{epsfig}
\usepackage{hyperref}
\usepackage{bbold}
\usepackage[normalem]{ulem}
\usepackage{cleveref}
\usepackage{footmisc}

\newcommand{\stkout}[1]{\ifmmode\text{\sout{\ensuremath{#1}}}\else\sout{#1}\fi}

\newcommand{\be}{\begin{equation}}
\newcommand{\ee}{\end{equation}}
\newcommand{\bea}{\begin{eqnarray}}
\newcommand{\eea}{\end{eqnarray}}

\def\a{\alpha}
\def\b{\beta}
\def\g{\gamma}
\def\G{\Gamma}
\def\d{\delta}

\def\th{\theta}
\def\Th{\Theta}

\def\l{\lambda}

\def\m{\mu}
\def\n{\nu}

\def\p{\pi}

\def\w{\omega}

\def\blr{{\mathbf r}}

\def\callA{\mbox{$\mathcal{A}$}}

\def\callG{\mbox{$\mathcal{G}$}}

\def\callI{\mbox{$\mathcal{I}$}}

\def\callM{\mbox{$\mathcal{M}$}}
\def\callN{\mbox{$\mathcal{N}$}}

\def\callT{\mbox{$\mathcal{T}$}}

\def\ra{\rightarrow}

\def\de{\partial}

\def\Tr{{\rm Tr}}

\def\1op{\hat{\mathbbm{1}}}
\def\nn{\nonumber}
\def\rL{{\rm L}}
\def\rR{{\rm R}}
\def\rT{{\rm T}}

\begin{document}

\title{Non-equilibrium spectral functions from multi-terminal steady-state density functional theory}

\author{Stefan Kurth}
\affiliation{Nano-Bio Spectroscopy Group and European Theoretical Spectroscopy
Facility (ETSF), Dpto. de F\'{i}sica de Materiales,
Universidad del Pa\'{i}s Vasco UPV/EHU, Av. Tolosa 72,
E-20018 San Sebasti\'{a}n, Spain}
\affiliation{IKERBASQUE, Basque Foundation for Science, Maria Diaz de Haro 3,
E-48013 Bilbao, Spain}
\affiliation{Donostia International Physics Center (DIPC), Paseo Manuel de
  Lardizabal 4, E-20018 San Sebasti\'{a}n, Spain}

\author{David Jacob}
\affiliation{Nano-Bio Spectroscopy Group and European Theoretical Spectroscopy
Facility (ETSF), Dpto. de F\'{i}sica de Materiales,
Universidad del Pa\'{i}s Vasco UPV/EHU, Av. Tolosa 72,
E-20018 San Sebasti\'{a}n, Spain}
\affiliation{IKERBASQUE, Basque Foundation for Science, Maria Diaz de Haro 3,
E-48013 Bilbao, Spain}

\author{Nahual Sobrino}
\affiliation{Donostia International Physics Center (DIPC), Paseo Manuel de
  Lardizabal 4, E-20018 San Sebasti\'{a}n, Spain}
\affiliation{Nano-Bio Spectroscopy Group and European Theoretical Spectroscopy
Facility (ETSF), Dpto. de F\'{i}sica de Materiales,
Universidad del Pa\'{i}s Vasco UPV/EHU, Av. Tolosa 72,
E-20018 San Sebasti\'{a}n, Spain}

\author{Gianluca Stefanucci}
\affiliation{Dipartimento di Fisica, Universit\`{a} di Roma Tor Vergata,
Via della Ricerca Scientifica 1, 00133 Rome, Italy; European Theoretical
Spectroscopy Facility (ETSF)}
\affiliation{INFN, Laboratori Nazionali di Frascati, Via E. Fermi 40,
00044 Frascati, Italy}

\begin{abstract}
  Multi-terminal transport setups allow to realize more complex measurements
  and functionalities (e.g., transistors) of nanoscale systems
  than the simple two-terminal arrangement.
  Here the steady-state density functional formalism (i-DFT) for the
  description of transport through nanoscale junctions with an arbitrary
  number of leads is developed. In a three-terminal setup and in the ideal
  STM limit where one of the electrodes (the ``STM tip'') is effectively
  decoupled from the junction, the formalism allows to extract its
  non-equilibrium spectral function (at arbitrary temperature) while a bias is
  applied between the other two electrodes. Multi-terminal i-DFT is shown to be
  capable of describing the splitting of the Kondo resonance in an Anderson
  impurity in the presence of an applied bias voltage, as predicted by
  numerically exact many-body approaches.
\end{abstract}

\maketitle


\section{Introduction}

Nanoscale or molecular junctions can now be made in the lab by connecting,
e.g., individual atoms, molecules or clusters to electrodes, which may become
the building blocks for prospective applications in Molecular Electronics
\cite{Cuniberti::2005,Cuevas::2010} and/or Quantum
Technologies\cite{Dowling:2003}. On a more fundamental level, nanoscale
junctions under an applied bias voltage are used to experimentally
\cite{Cuevas::2010,Ward:2011,Tewari:2018} study the many-body problem of
interacting electrons driven out of equilibrium, and
to probe nanoscale quantum systems by differential conductance
$dI/dV$ spectroscopy~\cite{Wiesendanger:book:1994,Hamers:chapter:2001,Madhavan:Science:1998,Hirjibehedin:Science:2007}. 

The description of transport through real nanoscale junctions presents a
major theoretical challenge since quantum effects, the atomistic details of
the junction, electronic interactions and the out-of-equilibrium situation
have to be properly taken into account. In its simplest form a nanoscale
junction can be described by a single impurity Anderson model (SIAM) coupled
to two leads at different chemical potentials that define the bias voltage
across the junction. An intriguing prediction for this model is the
splitting of the Kondo resonance under an applied bias
voltage~\cite{Meir:PRL:1993,Wingreen:PRB:1994,Sun:PRB:2001,Krawiec:PRB:2002,Shah:PRB:2006,Fritsch:PRB:2010,Cohen:PRL:2014}.
Experimentally, this effect cannot be seen directly in the $dI/dV$ of a two
terminal device, as it only shows up in the non-equilibrium spectral
function of the junction. However, it can be measured in a three-terminal
setup where one of the electrodes is very weakly coupled and serves as a
probe~\cite{Lebanon:PRB:2001,Sun:PRB:2001,DeFranceschi:PRL:2002,Leturcq:PRL:2005}.

In general, many-body techniques for solving the out-of-equilibrium problem are computationally too
demanding to be applied to more than relatively simple model systems such as the SIAM and slightly more
complex models.
Owing to its conceptual simplicity and computational efficiency, the now standard approach for realistic
modeling of electronic transport in nanoscale junctions
combines density functional theory (DFT) calculations with the Landauer-B\"uttiker approach (LB)
to transport~\cite{Taylor:PRB:2001a,Palacios:PRB:2002}.
While the LB-DFT approach properly takes into account atomistic details of the
junctions as well as quantum effects, it is formally incomplete in the sense
that there is no guarantee that it gives the correct current through the
interacting system even if the exact exchange-correlation functional is
used~\cite{Kurth:PRL:2013,Stefanucci:PSSB:2013}. 
It is therefore not surprising, that the LB-DFT formalism does not capture all
aspects of correlated electronic transport, namely Coulomb blockade and
Kondo physics, 
although under special circumstances some of these aspects may be
correctly described in a surprisingly simple manner
\cite{Stefanucci:PRL:2011,Bergfield:PRL:2012,Troester:PRB:2012}.
Combination of the LB-DFT approach with many-body methods incorporates 
electronic correlations (originating from a relatively small subspace) into
the description of electronic transport through realistic
systems~\cite{Jacob:PRL:2009,Jacob:PRB:2010,Jacob:JPCM:2015,Droghetti:PRB:2017},
but suffers from the infamous double-counting problem.

Recently, a novel approach, called steady-state DFT (or i-DFT), has been
devised to describe the steady-state transport through nanoscale junctions
driven out of equilibrium in a DFT
framework~\cite{Stefanucci:NL:2015,Kurth:PRB:2016,Kurth:JPCM:2017}.
In i-DFT the steady current through the nanoscale junction is an
additional fundamental ``density'' variable and the bias
voltage across the junction is the corresponding potential.
Provided that good approximations for the functionals are found, this approach
is able to describe the full phenomenology of electronic correlations in
transport through nanoscale junctions.
Moreover, the i-DFT formalism can be applied to extract the
{\em equilibrium} many-body spectral function from a DFT
calulation~\cite{Jacob:NL:2018}.

In this Letter we generalize i-DFT to the multi-terminal situation, and
then consider the specific situation of a junction connected to three
electrodes. We show how in the ``ideal STM setup'' where one of the
electrodes is only weakly coupled to the system, one can extract the
{\em non-equilibrium} many-body spectral function of the junction at
arbitrary temperature and bias between the other two electrodes. We apply the
approach to the SIAM for which we construct an approximate xc functional which
partially captures the splitting of the Kondo peak under finite bias. We
also identify the crucial feature of the xc functional needed to fully
describe the splitting of the Kondo peak.

\begin{figure}
\includegraphics[width=0.8\linewidth]{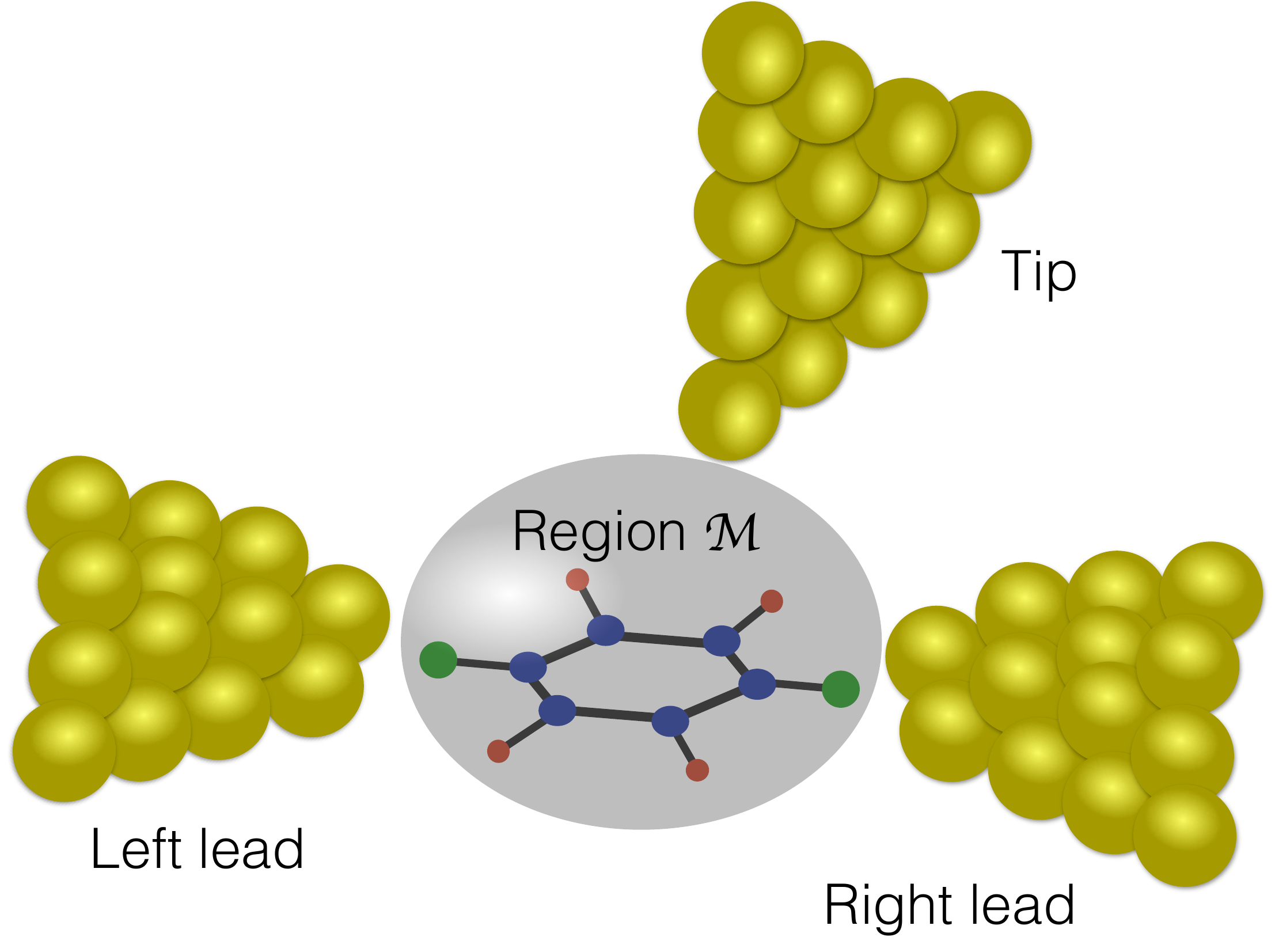}
\caption{\label{scheme}
Schematic drawing of a three-terminal nanoscale junction. A molecule $\callM$
is coupled to a left (L) and right (R) leads at voltages $V_{\rL}=-V_{\rR}=V/2$
with temperature $\Theta$ and to a tip (T) at voltage $V_{\rT}$ with zero
temperature $\Theta_{\rT}=0$. The molecule is contained in the region of
space $\callM$. }
\label{multiterm}    
\end{figure}

\section{Non-equilibrium spectral functions}

Here we breifly recall how to calculate out-of-equilibrium spectral functions
from transport measurements \cite{Meir:PRL:1993,Wingreen:PRB:1994,Sun:PRB:2001,Krawiec:PRB:2002,Shah:PRB:2006,Fritsch:PRB:2010,Cohen:PRL:2014}. We consider a 
three-terminal molecular junction as illustrated in Fig.~\ref{multiterm}.
 Two electrodes, the left 
($\a=\rL$) and right ($\a=\rR$) ones, have voltages 
$V_{\rL}=-V_{\rR}=V/2$ (gauge fixing) and the same finite temperature 
$\Theta_{\rL}=\Theta_{\rR}=\Theta$. The third 
electrode plays the role of a tip ($\a=\rT$) and is kept at zero 
temperature and voltage $V_{\rT}$. The contact between the 
tip and the nanoscopic region is described by the energy-independent
hybridization $\G_{\rT}$ whose indices run over a suitable one-electron orbital
basis for the considered molecule. The $\G_{\rT}$ matrix, aside from being 
constrained to be symmetric and positive 
semi-definite, will be varied at will.

According to Meir and Wingreen~\cite{Meir:PRL:1992} 
the current $I_{\rT}$ flowing out of the tip is given by (henceforth 
$\int\equiv \int\frac{d\w}{2\p}$)
\be
I_{\rT}=2\int\Tr\left[f_{\rT}(\w-V_{\rT})\G_{\rT}A(\w)+i\G_{\rT}G^{<}(\w)\right]
\label{Itip}
\ee
where $A(\w)=i\left[G^{>}(\w)-G^{<}(\w)\right]$ is the nonequilibrium,
finite-temperature many-body spectral function in terms of the lesser/greater 
Green's functions whereas $f_{\rT}(\w)=\th(-\w)$ is the zero-temperature Fermi 
function of the tip. In Eq.~(\ref{Itip}) the trace is over the  
indices of the molecular one-electron basis.
Similar to what we showed in previous work~\cite{Jacob:NL:2018} 
in the ideal Scanning Tunneling Microscopy (STM) limit, $\G_{\rT}\to 0$,
the Green's functions $G^{\lessgtr}$ are not affected by a change of the tip voltage and 
hence
\be
\lim_{\G_{\rT}\to 0}\frac{\de G^{\lessgtr}(\w)}{\de V_{\rT}}=0.
\label{STM1}
\ee
We then consider a hybridization of the form
\be
\G_{\rm 
T}=\g_{\rT}\bigg[\eta_{p}|p\rangle\langle p|+\eta_{q}|q\rangle\langle q|+\eta_{pq}
\big(|p\rangle\langle q|+|q\rangle\langle p|\big)\bigg]
\ee
This operator is symmetric and positive semi-definite for all 
$\g_{\rm T},\eta_{p},\eta_{q}>0$ and $|\eta_{pq}|\le\sqrt{\eta_{p}\eta_{q}}$.
Taking into account Eq.~(\ref{STM1}) it is straightforward to show that
\be
\lim_{\g_{\rT}\to 0}\frac{1}{\g_{\rT}}\frac{\de I_{\rT}}{\de 
V_{\rT}}=\frac{\callA(V_{\rT})}{\p}
\label{IversusA}
\ee
where
\be
\callA(\w)=\eta_{p}A_{pp}(\w)+\eta_{q}A_{qq}(\w)+\eta_{pq}
\big[A_{pq}(\w)+A_{qp}(\w)\big]
\label{callA}
\ee
is a linear combination of the matrix elements of the spectral function,
i.e. $A_{pq}(\w) = \langle p |A(\w) | q \rangle$. Choosing, e.g.,
$\eta_{p}=1$ and $\eta_{q}=0$ we can obtain all diagonal elements 
$A_{pp}=\callA$ by varying $p$. Subsequently we can extract the off-diagonal
elements $A_{pq}+A_{qp}=\callA-A_{pp}-A_{qq}$ by 
setting $\eta_{pq}=\eta_{p}=\eta_{q}=1$. 

\section{Multi-terminal i-DFT}

In Ref.~\cite{Jacob:NL:2018} we showed how to calculate equilibrium 
and zero-temperature spectral functions from the i-DFT 
approach~\cite{Stefanucci:NL:2015}.
For nonequilibrium and finite-temperature spectral functions we 
have to generalize i-DFT to multi-terminal setups, with electrodes 
at different voltages and temperatures. 

We consider a nanoscopic region $\callM$ containing a 
quantum dot or molecule and a number of electrodes $\a = 1,..,\callN$,
as depicted schematically in Fig.~\ref{multiterm} for $\callN=3$. 
The system is assumed to be in a steady state characterized by
temperatures $\Theta_{\a}$ and external voltages $V_{\a}$ in electrode $\a$
and by a gate voltage $v(\blr)$ in $\callM$.
As long as region $\callM$ is finite there are no constraints on the 
shape of its boundaries.
Due to gauge 
invariance the same steady-state is attained by shifting all voltages 
by a constant energy $W$, i.e., $V_{\a}\to V_{\a}+W$ and $v(\blr)\to 
v(\blr)+W$. Let $I_{\a}$ be the longitudinal current flowing out 
of electrode $\a$ and $n(\blr)$ be the density in the nanoscopic 
region. Due to charge conservation (consequence of the aforementioned gauge 
invariance) the currents  fulfill $\sum_{\a}I_{\a}=0$.
With a similar proof as the one published in 
Ref.~\cite{Stefanucci:NL:2015} we can state the multi-terminal 
generalization of the i-DFT theorem:
\\
{\em Theorem}: There exists a one-to-one 
map between the set of ``densities'' $(n,I_1,\ldots,I_{\tiny \callN})$ with $\sum_{\a}I_{\a}=0$ and the set of 
``potentials'' $(v,V_1,\ldots,V_{\tiny \callN})$
up to a constant shift $W$. The bijectivity of the 
map is guaranteed in a finite (and gate dependent) region around  
zero voltages $V_{\a}$ for any set of finite temperatures $\Theta_{\a}$.

According to the multi-terminal i-DFT theorem there exists a unique set 
of Kohn-Sham (KS) potentials $(v_{s}$, $V_{1,s},\ldots,V_{\tiny \callN,s})$
which in the 
noninteracting system reproduce the density $n(\blr)$ and  
currents $\callI=(I_1, \ldots, I_{\tiny \callN})$ of the interacting system  (here we are
assuming that the density and the currents are non-interacting 
representable). Following the 
KS procedure we define the exchange-correlation (xc) voltages $V_{\a,\rm 
xc}[n,\callI]=V_{\a,s}[n,\callI]-V_{\a}[n,\callI]$ and the Hartree-xc (Hxc) 
gate voltage $v_{\rm Hxc}[n,\callI]=v_{s}[n,\callI]-v[n,\callI]$ 
(which are functionals of the density in $\callM$ and the currents) and 
then calculate the interacting density and currents by solving
self-consistently the equations
\bea
n(\blr)\!&=&\!2\sum_{\a}\!\int f_{\a}(\w-V_{\a}-V_{\a,\rm xc}[n,\callI])\,
A_{\a,s}(\w,\blr),\quad
\\
I_{\a}\!&=&\! 2\sum_{\a'}\!\int
\big[f_{\a}(\w-V_{\a}-V_{\a,\rm xc}[n,\callI])
\nn\\
&-&
f_{\a'}(\w-V_{\a'}-V_{\a',\rm 
xc}[n,\callI])\big]\,\callT_{\a\a',s}(\w),
\eea
where $f_{\a}(\w)=1/(e^{\w/\Theta_{\a}}+1)$ is the Fermi function of lead 
$\a$ at temperature $\Theta_{\a}$.
In the KS equations $A_{\a,s}(\w,\blr)=\langle\blr| G^{\rm R}_{s}(\w)\G_{\a}(\w)G^{\rm 
A}_{s}(\w)|\blr\rangle$ is the partial KS spectral function written 
in terms of the retarded/advanced KS Green's functions $G^{\rm 
R/A}_{s}$ and hybridization
$\G_{\a}(\w)$  due to lead $\a$, whereas 
$\callT_{\a\a',s}(\w)=\Tr\left[ G^{\rm R}_{s}(\w)\G_{\a}(\w)G^{\rm 
A}_{s}(\w)\G_{\a'}(\w)\right]$ are the KS transmission probabilities.

\section{Spectral function from i-DFT}

We specialize the multi-terminal i-DFT theorem to the three-terminal 
case previously discussed. Let us fix the gauge according to 
$V_{{\rm L},s}=-V_{{\rm R},s}=V_{s}/2$ and let us consider 
the combination $I=(I_{L}-I_{R})/2$ and  $I_{\rT}$ as the two 
independent currents. Then the triple $v_{\rm Hxc}=v_{\rm Hxc}[n,I_{\rT},I]$, 
$V_{\rT,\rm xc}=V_{\rT,\rm xc}[n,I_{\rT},I]$ and $V_{\rm xc}=V_{\rm xc}[n,I_{\rT},I]$ are 
functionals  of the triple $n$, $I_{\rT}$ and $I$ (here $V_{\rm 
xc}[n,I_{\rT},I]=V_{s}[n,I_{\rT},I]-V[n,I_{\rT},I]$).
Considering $n,I_{\rT}$ and $I$ as interacting functionals of the physical 
voltages $v,V_{\rT}$ and $V$, Eq.~(\ref{STM1}) implies that
$\partial{n(\blr)}/\partial{V_\rT}\ra0$ and $\partial{I}/\partial{V_\rT}\ra0$
for $\Gamma_\rT\ra0$, and by the chain rule it thus follows that 
\be
\lim_{\G_{\rT}\to 0}\frac{\de v_{\rm Hxc}}{\de V_{\rT}}=
\lim_{\G_{\rT}\to 0}\frac{\de V_{\rm xc}}{\de V_{\rT}}=0.
\label{STM2}
\ee
In the same spirit as in our previous work~\cite{Jacob:NL:2018}, 
we now take advantage of these relations in order to express the spectral
function $A$ in terms of the KS spectral function $A_{s}$. In the 
noninteracting KS system the tip current
is given by Eq. (1), replacing $A(\w)$ with the KS spectral
function $A_{s}=\sum_{\a}A_{\a,s}$ and $G^<(\w)$ by the KS lesser GF $G^{<}_{s}$. 
Taking into account  Eq.~(\ref{STM2}) and the fact that $\G_{\rT}$ is 
energy-independent we find
\be
\lim_{\g_{\rT}\to 0}\frac{1}{\g_{\rT}}\frac{\de I_{\rT}}{\de 
V_{\rT}}=\frac{\callA_{s}(V_{\rT}+V_{\rT,\rm xc})}{\p}
\left(1+\frac{\de V_{\rT,\rm xc}}{\de I_{\rT}}\frac{\de I_{\rT}}{\de 
V_{\rT}}\right)
\ee
where $\callA_{s}$ is defined as in Eq.~(\ref{callA}) with $A\to
A_{s}$. Combining this result with Eq.~(\ref{IversusA}) we arrive at the 
first main result of this work 
\be
\callA(\w)=\lim_{\g_{\rT}\to 0}\frac{\callA_{s}(\w+V_{\rT,\rm xc}(\w))}
{1-\frac{\g_{\rT}}{\p}\frac{\de V_{\rT,\rm xc}(\w)}{\de 
I_{\rT}}\callA_{s}(\w+V_{\rT,\rm xc}(\w))},
\label{main1}
\ee
which generalizes the corresponding result of Ref.~\cite{Jacob:NL:2018}
to nonequilibrium spectral functions.
Here we have made explicit the dependence of $V_{\rT,\rm xc}$ on 
$\w=V_{\rT}$ through its dependence on $I_{\rT}$.
Choosing, e.g., $\eta_{p}=1$ and $\eta_{q}=0$, 
Eq.~(\ref{main1}) provides a relation between $A_{pp}$ and 
$A_{s,pp}$. The off-diagonal combination $A_{pq}+A_{qp}$ does instead 
follow by setting $\eta_{pq}=\eta_{p}=\eta_{q}=1$. 
We also observe that both $\callA_s$ and $\callA$ are normalized to the
same value, i.e. $\int\callA(\w)=\int \callA_{s}(\w)$ as it should
be~\footnote{
    This follows by integrating over $\w$ both sides of Eq.~(\ref{main1}),
    changing variable $\w'=\w+V_{\rT,\rm xc}(\w)$ in the r.h.s.
    and taking into account 
    the Jacobian 
    $\frac{{\rm d} \w'}{{\rm d} \w}=
    1/ (1 - \frac{\g_{\rT}}{\pi} 
    \frac{\partial V_{\rm T,xc}(\w)}{\partial I_T}\callA_s(\w +V_{\rm T,xc}(\w)))$.
}.

\section{i-DFT potentials for the Anderson model}
We apply the i-DFT  framework to the single-impurity Anderson model 
(SIAM) with charging energy $U$. Since the SIAM nanoscopic region has only one electronic 
degree of freedom the 
density $n=N$ coincides with the impurity occupation $N$, and 
all hybridization matrices are scalar. We then  write 
$\G_{\rm T}=\g_{\rT}$ for the tip and consider energy-independent  
left/right hybridizations  $\g_{\rL/\rR}$. 
The i-DFT self-consistent equations for $N$, $I_{\rT}$ and $I$ read
\be
N=2\int\sum_{\a=\rL,\rR,\rT}
\tilde{f}_{\a}(\w)
\frac{\g_{\a}}{\g}
A_{s}(\w)
\ee
\be
I_{\rT}=2\g_\rT \!\int \!
\left[ \frac{\g_{\rL}+\g_{\rR}}{\g}\tilde{f}_{\rT}(\w)
-\!\!\sum_{\a=\rL,\rR} \frac{\g_{\a}}{\g} \tilde{f}_{\a}(\w)
\right]\!\!
A_{s}(\w)
\ee
\bea
I\!=\!2\int\left[
\g_{\rL}\frac{2\g_{\rR}+\g_{\rT}}{\g}\tilde{f}_{\rL}(\w)-
\g_{\rR}\frac{2\g_{\rL}+\g_{\rT}}{\g}\tilde{f}_{\rR}(\w)
\right.
\nn\\
\left.
+\frac{\g_{\rT}(\g_{\rL}-\g_{\rR})}{\g}\tilde{f}_{\rT}(\w)
\right]\!A_{s}(\w)\quad
\eea
where we have defined $\tilde{f}_{\a}(\w)\equiv 
f_{\a}(\w-V_{\a}-V_{\a,\rm xc})$ as the 
shifted Fermi function and $\g\equiv \g_{\rL}+\g_{\rR}+\g_{\rT}$. The 
KS spectral function is simply $A_{s}(\w)=\ell_{\g}(\w-v-v_{\rm Hxc})$ with 
the Lorentzian $\ell_{\g}(\w)= \g/(\w^{2}+\g^{2}/4)$.

In order to derive an approximation for the i-DFT potentials we 
observe that in the interacting system the current flowing out of lead $\a$ 
reads $I_{\a}= 
2\int\left[f_{\a}(\w-V_{\a})\g_{\a}A(\w)+i\g_{\a}G^{<}(\w)
\right]$. 
Taking into account that the impurity 
occupation is $N=-2i\int\frac{d\w}{2\p} G^{<}(\w)$ we get
\be
N+\frac{I_{\a}}{\g_{\a}}= 2 \int f_{\a}(\w-V_{\a}) A(\w).
\label{ia2}
\ee
In the CB regime, i.e., for temperatures $\Theta_{\rL}=\Theta_{\rR}=\Theta$
larger than the Kondo temperature (at ph symmetry
\cite{JakobsPletyukhovSchoeller:10}) $\Theta_{K}
= \frac{4}{\pi}\sqrt{U \g} 
\exp\left( 
-\frac{\pi}{4}\left(\frac{U}{\g}-\frac{\g}{U}\right)\right)$ but smaller than
$\g_{\rL/\rR}$, the interacting spectral function is well approximated 
by \cite{Stefanucci:NL:2015,dittmann2017non}
\be
A(\w) = \frac{N}{2} l_{\g}(\w-v-U) + \left(1-\frac{N}{2}\right) 
l_{\g}(\w-v).
\label{model_specfunc}
\ee
Inserting 
Eq.~(\ref{model_specfunc}) in the r.h.s. of Eq.~(\ref{ia2}) we get 
the same expression obtained in Ref.~\cite{Stefanucci:NL:2015} for the 
two-terminal set-up. 
Therefore, the CB
reverse-engineered xc potentials can be parametrized in the same 
manner
\be
v^{\rm CB}_{\rm Hxc} - V^{\rm CB}_{\a,\rm xc}
\approx \frac{U}{2} + \frac{U}{\pi} {\rm atan} \left[
\frac{N+I_{\a}/(2\g_{\a})-1}{\nu W(\Theta_{\a})} \right]
\label{xcpot}  
\ee
with $\nu=1$, $W(\Theta)=0.16\times(\g/U)(1+9(\Theta/\g)^{2})$ and
$I_{\rL}=I-I_{\rT}/2$, 
$I_{\rR}=-I-I_{\rT}/2$ (as follows from charge conservation). 
From Eqs.~(\ref{xcpot}) we can easily extract an explicit form of the 
(H)xc potentials $v^{\rm CB}_{\rm Hxc}$, $V^{\rm CB}_{\rT,\rm xc}$ and $V^{\rm CB}_{\rm 
xc}=2V^{\rm CB}_{\rL,\rm xc}=-2V^{\rm CB}_{\rR,\rm xc}$ in terms of $N,I_{\rT}$ and $I$.

The (H)xc potentials in Eq.~(\ref{xcpot}) are certainly inadequate 
for temperatures $\Theta\lesssim \Theta_{K}$. In particular for
$\Theta=0$ the Friedel sum rule implies that the zero-bias interacting and KS 
conductances $\callG_{\a\b}=\de I_{\a}/\de V_{\b}$ and $\callG_{s,\a\b}=\de 
I_{\a}/\de V_{s,\b}$ are identical
\footnote{Using the Friedel sum-rule one can show that 
 $\callG_{\a\b}=C_{\a\b}(\g_\rT,\g_\rL,\g_\rR)A(\m)$ where the 
 prefactor $C_{\a\b}$ depends only on the hybridizations and 
 $A(\m)$ is the interacting spectral function at chemical 
 potential $\m$ (which is set to zero in our case). Since 
 $A(\m)=\frac{4}{\g}\sin^2(\p N/2)$ and since in i-DFT the KS 
 occupation $N$ is the same as the interacting $N$ we conclude 
 that the interacting and KS conductances are the same.}.
Since (repeated 
indices are summed over)
\bea
\callG_{\a\b}=\frac{\de I_{\a}}{\de V_{s,\m}}
\frac{\de V_{s,\m}}{\de V_{\b}}
=
\callG_{s,\a\m}\left(\d_{\m\b}+
\frac{\de V_{\m,\rm xc}}{\de I_{\n}}\callG_{\n\b}\right)
\eea
the zero-temperature xc voltages must fulfill
$\de V_{\m,\rm xc}/\de I_{\n}=0$ at zero currents. 
We incorporate this property in $v_{\rm Hxc}$ and $V_{\rm xc}$ using 
the same parametrization proposed in 
Ref.~\cite{Stefanucci:NL:2015} for the two-terminal case, i.e., 
for $\g_{\rT}=0$, which has been shown to be accurate in a wide range of 
temperatures and charging energy. For  
$V_{\rT,\rm xc}$ we propose 
\be
V_{\rT,\rm xc}(N,I_{\rT},I)= \big[
1-b(N)a_{\rT}(I_{\rT})a(I)\big]
V^{\rm CB}_{\rT,\rm xc}(N,I_{\rT},I)
\label{xc-tip-bias}
\ee
where in $V^{\rm CB}_{\rT,\rm xc}$ we now take $\nu=2$ \cite{Kurth:PRB:2016} and 
the functions $a_{\rT}$ and $a$ are similar to the one used in 
Ref.~\cite{Jacob:NL:2018} and read
\be
a_{\rT}(I_{\rT})=1-\frac{2}{\p}{\rm atan}\left[\l\left(
\frac{I_{\rT}}{W(0)\g_{\rT,\rm eff}}\right)^{2}\right]
\ee
\be
a(I)=1-\frac{2}{\p}{\rm atan}\left[\l\left(
\frac{I}{W(\Theta)\g_{\rm eff}}\right)^{2}\right]
\ee
with 
$\g_{\rT,\rm eff}=\frac{4\g_{\rT}(\g_{\rL}+\g_{\rR})}{\g}$, 
$\g_{\rm eff}=\frac{4\g_{\rL}\g_{\rR}}{\g_{\rL}+\g_{\rR}}$
and $\l = 0.16$ the same fit parameter  used in 
Ref.~\cite{Jacob:NL:2018}. For $b(N)$ we implement 
the same function as in Ref.~\cite{Stefanucci:NL:2015} but we replace 
the {\em two-terminal} conductance $\callG_{\rm univ}=dI/dV$ at the
ph symmetric gate $v=-U/2$, voltage $V=0$ and symmetric coupling
$\g_{\rL}=\g_{\rR}$ (this is a universal function depending 
only on the ratio $\Theta/\Theta_{K}$) with  
the three-terminal conductance $\callG_{\rT}=dI_{\rT}/dV_{\rT}$ at
the ph symmetric gate and voltages $V=V_{\rT}=0$:
\be
b(N=1)=1+\frac{1}{\left.\frac{\de V^{\rm CB}_{\rT,\rm xc}}{\de 
I_{\rT}}\right|_{\substack{N=1\\I=I_{\rT}=0}}}\left(
\frac{1}{\callG_{\rT}}-\frac{1}{\callG_{s,\rT}}\right).
\label{prova3}
\ee
One can show that 
$\callG_{\rT}=\frac{4\g_{\rT}(\g_{\rL}+\g_{\rR})}{\g^{2}}
\callG_{\rm univ}$. In Eq.~(\ref{prova3}) 
$\callG_{s,\rT}=dI_{\rT}/dV_{s,\rT}$ is the KS 
conductance at the same external potentials, i.e., ph gate and zero 
voltages.

\begin{figure}[t]
  \includegraphics[width=0.47\textwidth]{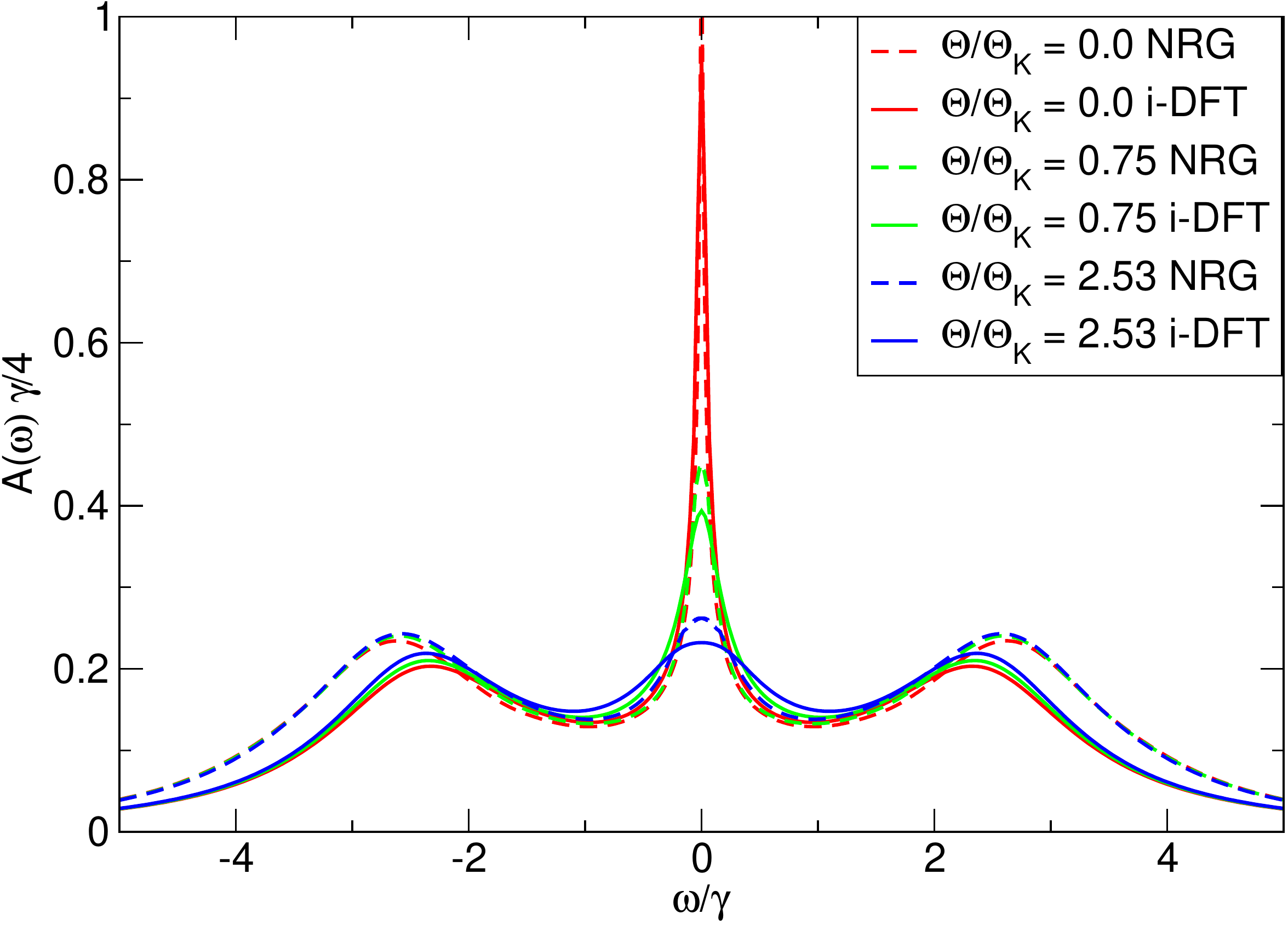}
  \caption{Equilibrium i-DFT spectral functions $A(\w)$ of the SIAM
    at ph symmetry for $U/\g=5$ for various
    temperatures compared with NRG results
    \cite{Motahari:PhD:2017,Requist:private:2017}. The Kondo temperature is
  $\Theta_{K}/\g\approx 0.066$. }
\label{compare_FT_NRG}
\end{figure}

\section{Results} 
As a first test we use our three-terminal i-DFT setup to compute 
the spectral function of the SIAM in thermal equilibrium for which
we can compare with results from numerical renormalization
group (NRG) techniques \cite{Motahari:PhD:2017,Requist:private:2017}, see
Fig.~\ref{compare_FT_NRG}. 
The i-DFT spectra agree reasonably well with the NRG ones although 
the height of the Kondo peak is slightly overestimated and 
for $\Th/\Th_{K} \gtrsim 2.5$
the Coulomb blockade side peaks are a bit too narrow.
In general, the finite temperature i-DFT spectra are of comparable quality as
the zero-temperature ones \cite{Jacob:NL:2018}. 

\begin{figure}[t]
  \includegraphics[width=0.46\textwidth]{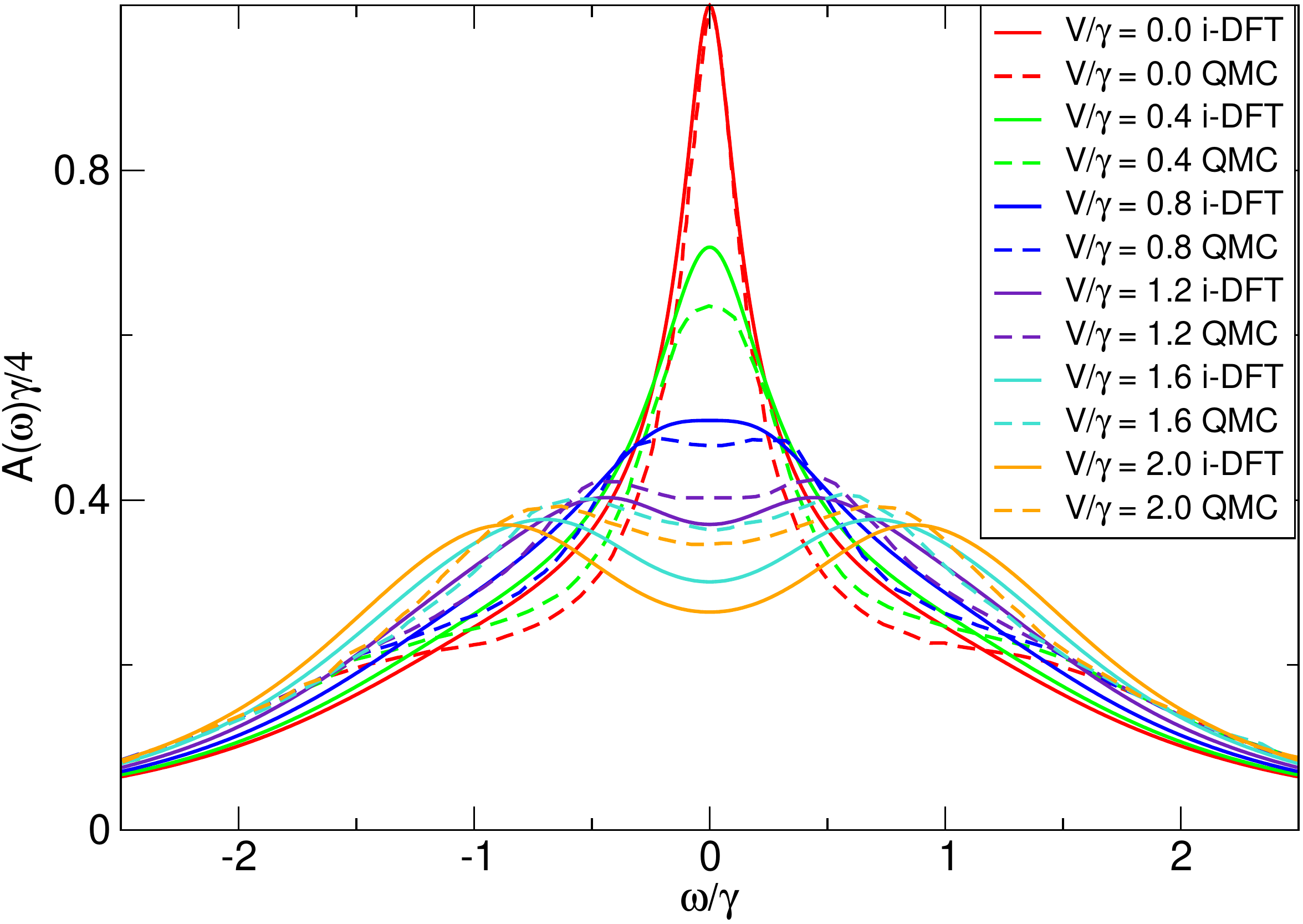}
  \caption{Comparison of i-DFT and QMC non-equilibrium spectral 
  functions from  Ref.~\cite{Bertrand:arxiv:2019} at
  particle-hole symmetry for $U/\g=2.5$ and zero temperature.
  The Kondo temperature is $\Theta_{K}/\g\approx 0.39$. }
\label{spec_comp_qmc}
\end{figure}

We now consider the zero-temperature, non-equilibrium SIAM and benchmark 
the i-DFT spectra against recent results from the Quantum Monte Carlo (QMC) 
approach~\cite{Bertrand:arxiv:2019}, see 
Fig.~\ref{spec_comp_qmc}. i-DFT reproduces all main qualitative features
of the QMC spectra. In particular, our simple functional of Eq.~(\ref{xc-tip-bias})
for the xc tip bias
is able to capture the finite-bias splitting of the Kondo peak
in this moderately correlated case $U/\g=2.5$. Nevertheless, in i-DFT
the splitting appears at somewhat higher biases 
and the distance between the peaks increases with bias faster 
than in QMC. We have done calculations for the same set of biases but 
at a finite temperature $\Th/\Th_{K}=0.6$ and observed no dramatic 
changes except for the suppression of the Kondo peak already at zero 
voltage.

\begin{figure}[t]
  \includegraphics[width=0.47\textwidth]{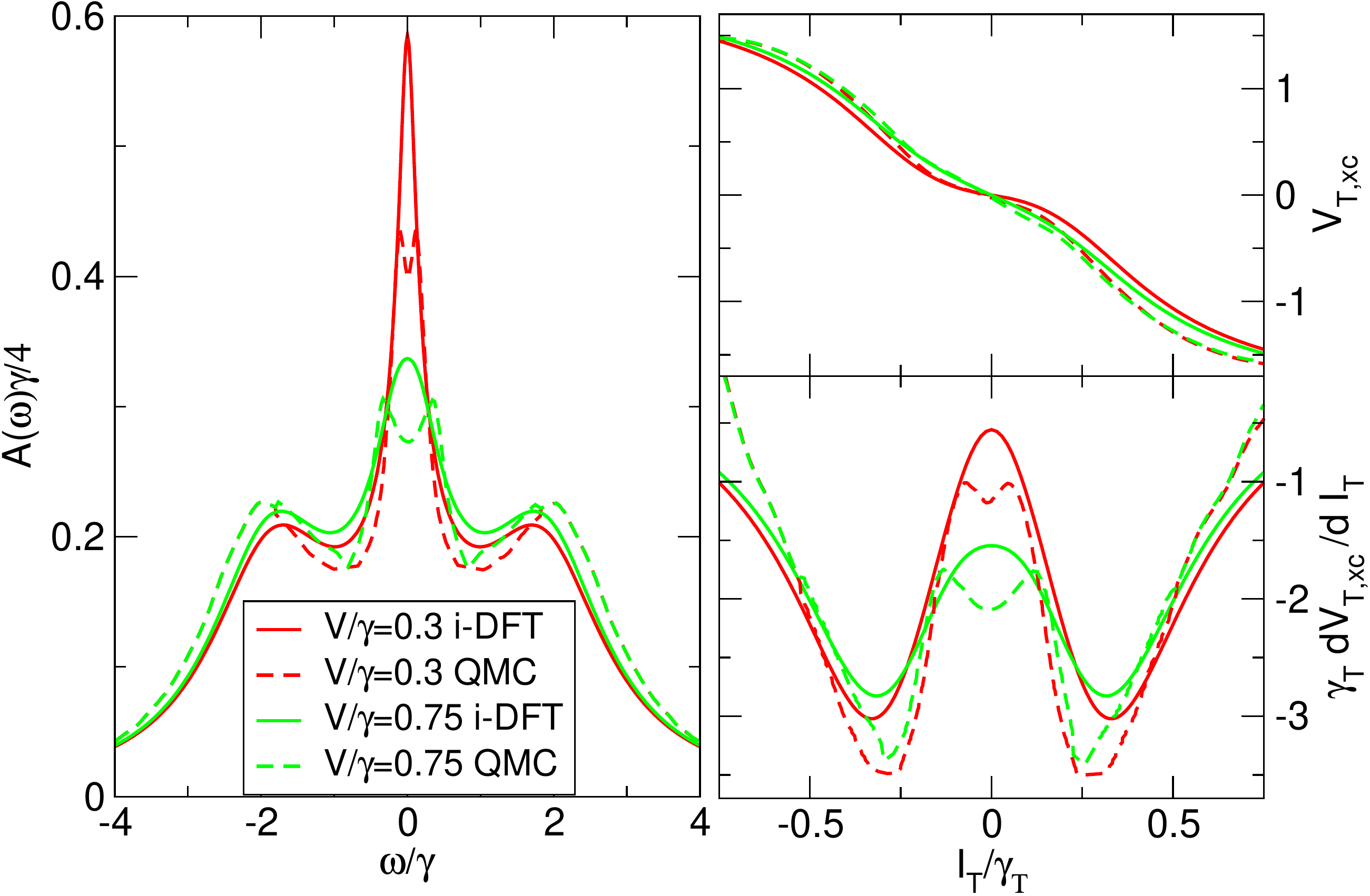}
  \caption{Left panel: i-DFT and QMC non-equilibrium spectral functions at
    particle-hole symmetry for $U/\g=4$ with QMC results from
    Ref.~\cite{Bertrand:arxiv:2019}. Upper right panel: xc tip bias as
    function of tip current $I_{\rm T}$ at $N=1$ and fixed current $I$ 
    corresponding to the two bias values. i-DFT results from our model tip xc
    bias of Eq.~(\ref{xc-tip-bias}), QMC results from reverse engineering using
    the QMC spectral function, see text. Lower right panel: derivatives of 
    $V_{\rm T,xc}$ of upper right panel w.r.t. $I_{\rm T}$.}
\label{spec_xcbias}
\end{figure}

In Fig.~\ref{spec_xcbias} (left panel) we compare i-DFT with QMC
non-equilibrium spectral functions \cite{Bertrand:arxiv:2019} for
a stronger interaction strength $U/\g=4$. 
Clearly our approximation to $V_{\rm T,xc}$ is missing a crucial 
feature since the Kondo splitting is totally absent in i-DFT. 
Below we highlight an exact property that 
$V_{\rm T,xc}$ must fulfill in order to capture the finite-bias splitting. 
The interacting spectral function in Eq.~(\ref{main1}) can also be written 
as
\be
A(\w) = \frac{\rm d}{{\rm d}\w} \int^{\w + V_{\rm T,xc}(\w)} 
{\rm d} \w' A_s(\w').
\label{spec_rev_eng}
\ee
Therefore, given a many-body (e.g., QMC) spectral function $A(\w)$, by integration of 
Eq.~(\ref{spec_rev_eng}) one can reverse-engineer the xc tip bias
$V_{\rm T,xc}$ which corresponds to the given $A$. In the upper right panel
of Fig.~\ref{spec_comp_qmc} we extracted $V_{\rm T,xc}$ as
function of $I_{\rm T}$ (for fixed values of $N$ and $I$)
corresponding to the QMC spectral functions of the left
panel of the same figure and compare to our i-DFT functional of
Eq.~(\ref{xc-tip-bias}). Although some differences are visible
our approximate xc tip bias seems to agree rather well with the 
reverse engineered one. The missing feature becomes evident
if we compare the derivatives of $V_{\rm T,xc}$ w.r.t. $I_{\rm 
T}$, see
lower right panel of Fig.~\ref{spec_comp_qmc}. 
While the derivative of the reverse engineered $V_{\rm T,xc}$ 
exhibits a double peak 
in the vicinity of $I_{\rm T}/\g_T\approx 0$, our approximation exhibits
only a single maximum at $I_{\rm T}/\g_T = 0$. 
Of course,
the height as well as the positions of the maxima depend on the current
$I$ between the left and right leads. 
We have verified that using the reverse engineered $V_{\rm T,xc}$
in Eq.~(\ref{main1}) the i-DFT and QMC spectral functions become 
indistinguishable.
The correct incorporation of the double peak feature
into an improved approximation for $V_{\rm T,xc}$ is 
beyond the scope of this work. However, the established existence of this 
xc bias constitues a proof-of-concept: i-DFT provides a 
numerically cheap method to calculate non-equilibrium spectral functions 
at zero and finite temperature.

\section{Conclusions}
We have generalized the i-DFT formalism for 
steady state transport through nanoscale junctions to the situation of
multiple electrodes. In particular, for a three-terminal setup in the
limit of vanishing coupling to one of the electrodes (ideal STM limit) we
have shown how to extract the {\em non-equilibrium} spectral function of the
junction at both zero and finite temperature extending earlier work
\cite{Jacob:NL:2018} which was restricted both to equilibrium and zero
temperature. For the specific situation of an Anderson model coupled to three
electrodes we have constructed an approximate xc functional by a relatively
simple “natural” generalization of already existing i-DFT functionals. This
approximation describes, at least for not too strong interactions, the
splitting of the Kondo peak at finite bias and yields results in reasonable
qualitative agreement with computationally more demanding many-body approaches
such as NRG and non-equilibrium QMC. Although for stronger interactions our
approximation does not capture the splitting of the Kondo peak, we were
nevertheless able to identify the missing feature which needs to be
incorporated in future improved functionals. In order to construct such
functionals, reliable reference results from other many-body methods are
certainly very welcome
\cite{Cohen:PRL:2014,Bertrand:arxiv:2019,Krivenko:arxiv:2019}.
However, once such approximations are available for a relatively simple
system such as the Anderson model, generalizations to more complicated
model systems (such as, e.g., multi-level systems) may actually be
relatively straightforward
\cite{Stefanucci:NL:2015,Kurth:JPCM:2017,Jacob:NL:2018,Kurth:EPJB:2018}. 
Since multi-terminal i-DFT is comparable in computational
effort to standard LB-DFT calculations, it is therefore suitable to study
systems currently inaccessible for accurate out-of-equilibrium many-body
methods.

S.K. acknowledges funding through a grant of the
"Ministerio de Economia y Competividad (MINECO)" (FIS2016-79464-P).
GS acknowledges EC funding through the RISE Co-ExAN (Grant No. GA644076)
and Tor Vergata University for financial support through the Mission
Sustainability Project 2DUTOPI.


\begin{thebibliography}{44}%
\makeatletter
\providecommand \@ifxundefined [1]{%
 \@ifx{#1\undefined}
}%
\providecommand \@ifnum [1]{%
 \ifnum #1\expandafter \@firstoftwo
 \else \expandafter \@secondoftwo
 \fi
}%
\providecommand \@ifx [1]{%
 \ifx #1\expandafter \@firstoftwo
 \else \expandafter \@secondoftwo
 \fi
}%
\providecommand \natexlab [1]{#1}%
\providecommand \enquote  [1]{``#1''}%
\providecommand \bibnamefont  [1]{#1}%
\providecommand \bibfnamefont [1]{#1}%
\providecommand \citenamefont [1]{#1}%
\providecommand \href@noop [0]{\@secondoftwo}%
\providecommand \href [0]{\begingroup \@sanitize@url \@href}%
\providecommand \@href[1]{\@@startlink{#1}\@@href}%
\providecommand \@@href[1]{\endgroup#1\@@endlink}%
\providecommand \@sanitize@url [0]{\catcode `\\12\catcode `\$12\catcode
  `\&12\catcode `\#12\catcode `\^12\catcode `\_12\catcode `\%12\relax}%
\providecommand \@@startlink[1]{}%
\providecommand \@@endlink[0]{}%
\providecommand \url  [0]{\begingroup\@sanitize@url \@url }%
\providecommand \@url [1]{\endgroup\@href {#1}{\urlprefix }}%
\providecommand \urlprefix  [0]{URL }%
\providecommand \Eprint [0]{\href }%
\providecommand \doibase [0]{http://dx.doi.org/}%
\providecommand \selectlanguage [0]{\@gobble}%
\providecommand \bibinfo  [0]{\@secondoftwo}%
\providecommand \bibfield  [0]{\@secondoftwo}%
\providecommand \translation [1]{[#1]}%
\providecommand \BibitemOpen [0]{}%
\providecommand \bibitemStop [0]{}%
\providecommand \bibitemNoStop [0]{.\EOS\space}%
\providecommand \EOS [0]{\spacefactor3000\relax}%
\providecommand \BibitemShut  [1]{\csname bibitem#1\endcsname}%
\let\auto@bib@innerbib\@empty
\bibitem [{\citenamefont {Cuniberti}\ \emph {et~al.}(2005)\citenamefont
  {Cuniberti}, \citenamefont {Fagas},\ and\ \citenamefont
  {Richter}}]{Cuniberti::2005}%
  \BibitemOpen
  \bibfield  {author} {\bibinfo {author} {\bibfnamefont {G.}~\bibnamefont
  {Cuniberti}}, \bibinfo {author} {\bibfnamefont {G.}~\bibnamefont {Fagas}}, \
  and\ \bibinfo {author} {\bibfnamefont {K.}~\bibnamefont {Richter}},\
  }\href@noop {} {\emph {\bibinfo {title} {Introducing Molecular
  Electronics}}}\ (\bibinfo  {publisher} {Springer, Berlin},\ \bibinfo {year}
  {2005})\BibitemShut {NoStop}%
\bibitem [{\citenamefont {Cuevas}\ and\ \citenamefont
  {Scheer}(2010)}]{Cuevas::2010}%
  \BibitemOpen
  \bibfield  {author} {\bibinfo {author} {\bibfnamefont {J.~C.}\ \bibnamefont
  {Cuevas}}\ and\ \bibinfo {author} {\bibfnamefont {E.}~\bibnamefont
  {Scheer}},\ }\href@noop {} {\emph {\bibinfo {title} {Molecular
  Electronics}}}\ (\bibinfo  {publisher} {World Scientific, Singapore},\
  \bibinfo {year} {2010})\BibitemShut {NoStop}%
\bibitem [{\citenamefont {J.P.Dowling}\ and\ \citenamefont
  {G.J.Milburn}(2003)}]{Dowling:2003}%
  \BibitemOpen
  \bibfield  {author} {\bibinfo {author} {\bibnamefont {J.P.Dowling}}\ and\
  \bibinfo {author} {\bibnamefont {G.J.Milburn}},\ }\href@noop {} {\bibfield
  {journal} {\bibinfo  {journal} {Phil. Trans. R. Soc. A}\ }\textbf {\bibinfo
  {volume} {361}},\ \bibinfo {pages} {3655} (\bibinfo {year}
  {2003})}\BibitemShut {NoStop}%
\bibitem [{\citenamefont {Ward}\ \emph {et~al.}(2011)\citenamefont {Ward},
  \citenamefont {Corley}, \citenamefont {Tour},\ and\ \citenamefont
  {Natelson}}]{Ward:2011}%
  \BibitemOpen
  \bibfield  {author} {\bibinfo {author} {\bibfnamefont {D.~R.}\ \bibnamefont
  {Ward}}, \bibinfo {author} {\bibfnamefont {D.~A.}\ \bibnamefont {Corley}},
  \bibinfo {author} {\bibfnamefont {J.~M.}\ \bibnamefont {Tour}}, \ and\
  \bibinfo {author} {\bibfnamefont {D.}~\bibnamefont {Natelson}},\ }\href@noop
  {} {\bibfield  {journal} {\bibinfo  {journal} {Nature Nanotech.}\ }\textbf
  {\bibinfo {volume} {6}},\ \bibinfo {pages} {33} (\bibinfo {year}
  {2011})}\BibitemShut {NoStop}%
\bibitem [{\citenamefont {Tewari}\ and\ \citenamefont {van
  Ruitenbeek}(2018)}]{Tewari:2018}%
  \BibitemOpen
  \bibfield  {author} {\bibinfo {author} {\bibfnamefont {S.}~\bibnamefont
  {Tewari}}\ and\ \bibinfo {author} {\bibfnamefont {J.}~\bibnamefont {van
  Ruitenbeek}},\ }\href@noop {} {\bibfield  {journal} {\bibinfo  {journal}
  {Nano Lett.}\ }\textbf {\bibinfo {volume} {18}},\ \bibinfo {pages} {5217}
  (\bibinfo {year} {2018})}\BibitemShut {NoStop}%
\bibitem [{\citenamefont {Wiesendanger}(1994)}]{Wiesendanger:book:1994}%
  \BibitemOpen
  \bibfield  {author} {\bibinfo {author} {\bibfnamefont {R.}~\bibnamefont
  {Wiesendanger}},\ }\href@noop {} {\emph {\bibinfo {title} {Scanning Probe
  Microscopy and Spectroscopy Methods and Applications}}}\ (\bibinfo
  {publisher} {Cambridge University Press},\ \bibinfo {address} {Cambridge},\
  \bibinfo {year} {1994})\BibitemShut {NoStop}%
\bibitem [{\citenamefont {Hamers}\ and\ \citenamefont
  {Padowitz}(2001)}]{Hamers:chapter:2001}%
  \BibitemOpen
  \bibfield  {author} {\bibinfo {author} {\bibfnamefont {R.~J.}\ \bibnamefont
  {Hamers}}\ and\ \bibinfo {author} {\bibfnamefont {D.~F.}\ \bibnamefont
  {Padowitz}},\ }in\ \href@noop {} {\emph {\bibinfo {booktitle} {Scanning Probe
  Microscopy and Spectroscopy: Theory, Techniques, and Applications}}},\
  \bibinfo {editor} {edited by\ \bibinfo {editor} {\bibfnamefont {D.~A.}\
  \bibnamefont {Bonnell}}}\ (\bibinfo  {publisher} {Wiley-VCH, Inc.},\ \bibinfo
  {address} {New York},\ \bibinfo {year} {2001})\BibitemShut {NoStop}%
\bibitem [{\citenamefont {Madhavan}\ \emph {et~al.}(1998)\citenamefont
  {Madhavan}, \citenamefont {Chen}, \citenamefont {Jamneala}, \citenamefont
  {Crommie},\ and\ \citenamefont {Wingreen}}]{Madhavan:Science:1998}%
  \BibitemOpen
  \bibfield  {author} {\bibinfo {author} {\bibfnamefont {V.}~\bibnamefont
  {Madhavan}}, \bibinfo {author} {\bibfnamefont {W.}~\bibnamefont {Chen}},
  \bibinfo {author} {\bibfnamefont {T.}~\bibnamefont {Jamneala}}, \bibinfo
  {author} {\bibfnamefont {M.~F.}\ \bibnamefont {Crommie}}, \ and\ \bibinfo
  {author} {\bibfnamefont {N.~S.}\ \bibnamefont {Wingreen}},\ }\href {\doibase
  10.1126/science.280.5363.567} {\bibfield  {journal} {\bibinfo  {journal}
  {Science}\ }\textbf {\bibinfo {volume} {280}},\ \bibinfo {pages} {567}
  (\bibinfo {year} {1998})}\BibitemShut {NoStop}%
\bibitem [{\citenamefont {Hirjibehedin}\ \emph {et~al.}(2007)\citenamefont
  {Hirjibehedin}, \citenamefont {Lin}, \citenamefont {Otte}, \citenamefont
  {Ternes}, \citenamefont {Lutz}, \citenamefont {Jones},\ and\ \citenamefont
  {Heinrich}}]{Hirjibehedin:Science:2007}%
  \BibitemOpen
  \bibfield  {author} {\bibinfo {author} {\bibfnamefont {C.~F.}\ \bibnamefont
  {Hirjibehedin}}, \bibinfo {author} {\bibfnamefont {C.-Y.}\ \bibnamefont
  {Lin}}, \bibinfo {author} {\bibfnamefont {A.~F.}\ \bibnamefont {Otte}},
  \bibinfo {author} {\bibfnamefont {M.}~\bibnamefont {Ternes}}, \bibinfo
  {author} {\bibfnamefont {C.~P.}\ \bibnamefont {Lutz}}, \bibinfo {author}
  {\bibfnamefont {B.~A.}\ \bibnamefont {Jones}}, \ and\ \bibinfo {author}
  {\bibfnamefont {A.~J.}\ \bibnamefont {Heinrich}},\ }\href {\doibase
  10.1126/science.1146110} {\bibfield  {journal} {\bibinfo  {journal}
  {Science}\ }\textbf {\bibinfo {volume} {317}},\ \bibinfo {pages}
  {1199–1203} (\bibinfo {year} {2007})}\BibitemShut {NoStop}%
\bibitem [{\citenamefont {Meir}\ \emph {et~al.}(1993)\citenamefont {Meir},
  \citenamefont {Wingreen},\ and\ \citenamefont {Lee}}]{Meir:PRL:1993}%
  \BibitemOpen
  \bibfield  {author} {\bibinfo {author} {\bibfnamefont {Y.}~\bibnamefont
  {Meir}}, \bibinfo {author} {\bibfnamefont {N.~S.}\ \bibnamefont {Wingreen}},
  \ and\ \bibinfo {author} {\bibfnamefont {P.~A.}\ \bibnamefont {Lee}},\
  }\href@noop {} {\bibfield  {journal} {\bibinfo  {journal} {Phys. Rev. Lett.}\
  }\textbf {\bibinfo {volume} {70}},\ \bibinfo {pages} {2601} (\bibinfo {year}
  {1993})}\BibitemShut {NoStop}%
\bibitem [{\citenamefont {Wingreen}\ and\ \citenamefont
  {Meir}(1994)}]{Wingreen:PRB:1994}%
  \BibitemOpen
  \bibfield  {author} {\bibinfo {author} {\bibfnamefont {N.~S.}\ \bibnamefont
  {Wingreen}}\ and\ \bibinfo {author} {\bibfnamefont {Y.}~\bibnamefont
  {Meir}},\ }\href {\doibase 10.1103/PhysRevB.49.11040} {\bibfield  {journal}
  {\bibinfo  {journal} {Phys. Rev. B}\ }\textbf {\bibinfo {volume} {49}},\
  \bibinfo {pages} {11040} (\bibinfo {year} {1994})}\BibitemShut {NoStop}%
\bibitem [{\citenamefont {Sun}\ and\ \citenamefont {Guo}(2001)}]{Sun:PRB:2001}%
  \BibitemOpen
  \bibfield  {author} {\bibinfo {author} {\bibfnamefont {Q.-f.}\ \bibnamefont
  {Sun}}\ and\ \bibinfo {author} {\bibfnamefont {H.}~\bibnamefont {Guo}},\
  }\href {\doibase 10.1103/PhysRevB.64.153306} {\bibfield  {journal} {\bibinfo
  {journal} {Phys. Rev. B}\ }\textbf {\bibinfo {volume} {64}},\ \bibinfo
  {pages} {153306} (\bibinfo {year} {2001})}\BibitemShut {NoStop}%
\bibitem [{\citenamefont {Krawiec}\ and\ \citenamefont
  {Wysoki\ifmmode~\acute{n}\else \'{n}\fi{}ski}(2002)}]{Krawiec:PRB:2002}%
  \BibitemOpen
  \bibfield  {author} {\bibinfo {author} {\bibfnamefont {M.}~\bibnamefont
  {Krawiec}}\ and\ \bibinfo {author} {\bibfnamefont {K.~I.}\ \bibnamefont
  {Wysoki\ifmmode~\acute{n}\else \'{n}\fi{}ski}},\ }\href {\doibase
  10.1103/PhysRevB.66.165408} {\bibfield  {journal} {\bibinfo  {journal} {Phys.
  Rev. B}\ }\textbf {\bibinfo {volume} {66}},\ \bibinfo {pages} {165408}
  (\bibinfo {year} {2002})}\BibitemShut {NoStop}%
\bibitem [{\citenamefont {Shah}\ and\ \citenamefont
  {Rosch}(2006)}]{Shah:PRB:2006}%
  \BibitemOpen
  \bibfield  {author} {\bibinfo {author} {\bibfnamefont {N.}~\bibnamefont
  {Shah}}\ and\ \bibinfo {author} {\bibfnamefont {A.}~\bibnamefont {Rosch}},\
  }\href {\doibase 10.1103/PhysRevB.73.081309} {\bibfield  {journal} {\bibinfo
  {journal} {Phys. Rev. B}\ }\textbf {\bibinfo {volume} {73}},\ \bibinfo
  {pages} {081309} (\bibinfo {year} {2006})}\BibitemShut {NoStop}%
\bibitem [{\citenamefont {Fritsch}\ and\ \citenamefont
  {Kehrein}(2010)}]{Fritsch:PRB:2010}%
  \BibitemOpen
  \bibfield  {author} {\bibinfo {author} {\bibfnamefont {P.}~\bibnamefont
  {Fritsch}}\ and\ \bibinfo {author} {\bibfnamefont {S.}~\bibnamefont
  {Kehrein}},\ }\href {\doibase 10.1103/PhysRevB.81.035113} {\bibfield
  {journal} {\bibinfo  {journal} {Phys. Rev. B}\ }\textbf {\bibinfo {volume}
  {81}},\ \bibinfo {pages} {035113} (\bibinfo {year} {2010})}\BibitemShut
  {NoStop}%
\bibitem [{\citenamefont {Cohen}\ \emph {et~al.}(2014)\citenamefont {Cohen},
  \citenamefont {Gull}, \citenamefont {Reichman},\ and\ \citenamefont
  {Millis}}]{Cohen:PRL:2014}%
  \BibitemOpen
  \bibfield  {author} {\bibinfo {author} {\bibfnamefont {G.}~\bibnamefont
  {Cohen}}, \bibinfo {author} {\bibfnamefont {E.}~\bibnamefont {Gull}},
  \bibinfo {author} {\bibfnamefont {D.~R.}\ \bibnamefont {Reichman}}, \ and\
  \bibinfo {author} {\bibfnamefont {A.~J.}\ \bibnamefont {Millis}},\
  }\href@noop {} {\bibfield  {journal} {\bibinfo  {journal} {Phys. Rev. Lett.}\
  }\textbf {\bibinfo {volume} {112}},\ \bibinfo {pages} {146802} (\bibinfo
  {year} {2014})}\BibitemShut {NoStop}%
\bibitem [{\citenamefont {Lebanon}\ and\ \citenamefont
  {Schiller}(2001)}]{Lebanon:PRB:2001}%
  \BibitemOpen
  \bibfield  {author} {\bibinfo {author} {\bibfnamefont {E.}~\bibnamefont
  {Lebanon}}\ and\ \bibinfo {author} {\bibfnamefont {A.}~\bibnamefont
  {Schiller}},\ }\href {\doibase 10.1103/PhysRevB.65.035308} {\bibfield
  {journal} {\bibinfo  {journal} {Phys. Rev. B}\ }\textbf {\bibinfo {volume}
  {65}},\ \bibinfo {pages} {035308} (\bibinfo {year} {2001})}\BibitemShut
  {NoStop}%
\bibitem [{\citenamefont {De~Franceschi}\ \emph {et~al.}(2002)\citenamefont
  {De~Franceschi}, \citenamefont {Hanson}, \citenamefont {van~der Wiel},
  \citenamefont {Elzerman}, \citenamefont {Wijpkema}, \citenamefont {Fujisawa},
  \citenamefont {Tarucha},\ and\ \citenamefont
  {Kouwenhoven}}]{DeFranceschi:PRL:2002}%
  \BibitemOpen
  \bibfield  {author} {\bibinfo {author} {\bibfnamefont {S.}~\bibnamefont
  {De~Franceschi}}, \bibinfo {author} {\bibfnamefont {R.}~\bibnamefont
  {Hanson}}, \bibinfo {author} {\bibfnamefont {W.~G.}\ \bibnamefont {van~der
  Wiel}}, \bibinfo {author} {\bibfnamefont {J.~M.}\ \bibnamefont {Elzerman}},
  \bibinfo {author} {\bibfnamefont {J.~J.}\ \bibnamefont {Wijpkema}}, \bibinfo
  {author} {\bibfnamefont {T.}~\bibnamefont {Fujisawa}}, \bibinfo {author}
  {\bibfnamefont {S.}~\bibnamefont {Tarucha}}, \ and\ \bibinfo {author}
  {\bibfnamefont {L.~P.}\ \bibnamefont {Kouwenhoven}},\ }\href {\doibase
  10.1103/PhysRevLett.89.156801} {\bibfield  {journal} {\bibinfo  {journal}
  {Phys. Rev. Lett.}\ }\textbf {\bibinfo {volume} {89}},\ \bibinfo {pages}
  {156801} (\bibinfo {year} {2002})}\BibitemShut {NoStop}%
\bibitem [{\citenamefont {Leturcq}\ \emph {et~al.}(2005)\citenamefont
  {Leturcq}, \citenamefont {Schmid}, \citenamefont {Ensslin}, \citenamefont
  {Meir}, \citenamefont {Driscoll},\ and\ \citenamefont
  {Gossard}}]{Leturcq:PRL:2005}%
  \BibitemOpen
  \bibfield  {author} {\bibinfo {author} {\bibfnamefont {R.}~\bibnamefont
  {Leturcq}}, \bibinfo {author} {\bibfnamefont {L.}~\bibnamefont {Schmid}},
  \bibinfo {author} {\bibfnamefont {K.}~\bibnamefont {Ensslin}}, \bibinfo
  {author} {\bibfnamefont {Y.}~\bibnamefont {Meir}}, \bibinfo {author}
  {\bibfnamefont {D.~C.}\ \bibnamefont {Driscoll}}, \ and\ \bibinfo {author}
  {\bibfnamefont {A.~C.}\ \bibnamefont {Gossard}},\ }\href {\doibase
  10.1103/PhysRevLett.95.126603} {\bibfield  {journal} {\bibinfo  {journal}
  {Phys. Rev. Lett.}\ }\textbf {\bibinfo {volume} {95}},\ \bibinfo {pages}
  {126603} (\bibinfo {year} {2005})}\BibitemShut {NoStop}%
\bibitem [{\citenamefont {Taylor}\ \emph {et~al.}(2001)\citenamefont {Taylor},
  \citenamefont {Guo},\ and\ \citenamefont {Wang}}]{Taylor:PRB:2001a}%
  \BibitemOpen
  \bibfield  {author} {\bibinfo {author} {\bibfnamefont {J.}~\bibnamefont
  {Taylor}}, \bibinfo {author} {\bibfnamefont {H.}~\bibnamefont {Guo}}, \ and\
  \bibinfo {author} {\bibfnamefont {J.}~\bibnamefont {Wang}},\ }\href@noop {}
  {\bibfield  {journal} {\bibinfo  {journal} {Phys. Rev. B}\ }\textbf {\bibinfo
  {volume} {63}},\ \bibinfo {pages} {245407} (\bibinfo {year}
  {2001})}\BibitemShut {NoStop}%
\bibitem [{\citenamefont {Palacios}\ \emph {et~al.}(2002)\citenamefont
  {Palacios}, \citenamefont {P\'erez-Jim\'enez}, \citenamefont {Louis},
  \citenamefont {SanFabi\'an},\ and\ \citenamefont
  {Verg\'es}}]{Palacios:PRB:2002}%
  \BibitemOpen
  \bibfield  {author} {\bibinfo {author} {\bibfnamefont {J.~J.}\ \bibnamefont
  {Palacios}}, \bibinfo {author} {\bibfnamefont {A.~J.}\ \bibnamefont
  {P\'erez-Jim\'enez}}, \bibinfo {author} {\bibfnamefont {E.}~\bibnamefont
  {Louis}}, \bibinfo {author} {\bibfnamefont {E.}~\bibnamefont {SanFabi\'an}},
  \ and\ \bibinfo {author} {\bibfnamefont {J.~A.}\ \bibnamefont {Verg\'es}},\
  }\href@noop {} {\bibfield  {journal} {\bibinfo  {journal} {Phys. Rev. B}\
  }\textbf {\bibinfo {volume} {66}},\ \bibinfo {pages} {035322} (\bibinfo
  {year} {2002})}\BibitemShut {NoStop}%
\bibitem [{\citenamefont {Kurth}\ and\ \citenamefont
  {Stefanucci}(2013)}]{Kurth:PRL:2013}%
  \BibitemOpen
  \bibfield  {author} {\bibinfo {author} {\bibfnamefont {S.}~\bibnamefont
  {Kurth}}\ and\ \bibinfo {author} {\bibfnamefont {G.}~\bibnamefont
  {Stefanucci}},\ }\href {\doibase 10.1103/PhysRevLett.111.030601} {\bibfield
  {journal} {\bibinfo  {journal} {Phys. Rev. Lett.}\ }\textbf {\bibinfo
  {volume} {111}},\ \bibinfo {pages} {030601} (\bibinfo {year}
  {2013})}\BibitemShut {NoStop}%
\bibitem [{\citenamefont {Stefanucci}\ and\ \citenamefont
  {Kurth}(2013)}]{Stefanucci:PSSB:2013}%
  \BibitemOpen
  \bibfield  {author} {\bibinfo {author} {\bibfnamefont {G.}~\bibnamefont
  {Stefanucci}}\ and\ \bibinfo {author} {\bibfnamefont {S.}~\bibnamefont
  {Kurth}},\ }\href {\doibase 10.1002/pssb.201349181} {\bibfield  {journal}
  {\bibinfo  {journal} {Physica Status Solidi (b)}\ }\textbf {\bibinfo {volume}
  {250}},\ \bibinfo {pages} {2378} (\bibinfo {year} {2013})}\BibitemShut
  {NoStop}%
\bibitem [{\citenamefont {Stefanucci}\ and\ \citenamefont
  {Kurth}(2011)}]{Stefanucci:PRL:2011}%
  \BibitemOpen
  \bibfield  {author} {\bibinfo {author} {\bibfnamefont {G.}~\bibnamefont
  {Stefanucci}}\ and\ \bibinfo {author} {\bibfnamefont {S.}~\bibnamefont
  {Kurth}},\ }\href {\doibase 10.1103/PhysRevLett.107.216401} {\bibfield
  {journal} {\bibinfo  {journal} {Phys. Rev. Lett.}\ }\textbf {\bibinfo
  {volume} {107}},\ \bibinfo {pages} {216401} (\bibinfo {year}
  {2011})}\BibitemShut {NoStop}%
\bibitem [{\citenamefont {Bergfield}\ \emph {et~al.}(2012)\citenamefont
  {Bergfield}, \citenamefont {Liu}, \citenamefont {Burke},\ and\ \citenamefont
  {Stafford}}]{Bergfield:PRL:2012}%
  \BibitemOpen
  \bibfield  {author} {\bibinfo {author} {\bibfnamefont {J.~P.}\ \bibnamefont
  {Bergfield}}, \bibinfo {author} {\bibfnamefont {Z.-F.}\ \bibnamefont {Liu}},
  \bibinfo {author} {\bibfnamefont {K.}~\bibnamefont {Burke}}, \ and\ \bibinfo
  {author} {\bibfnamefont {C.~A.}\ \bibnamefont {Stafford}},\ }\href {\doibase
  10.1103/PhysRevLett.108.066801} {\bibfield  {journal} {\bibinfo  {journal}
  {Phys. Rev. Lett.}\ }\textbf {\bibinfo {volume} {108}},\ \bibinfo {pages}
  {066801} (\bibinfo {year} {2012})}\BibitemShut {NoStop}%
\bibitem [{\citenamefont {Tr\"oster}\ \emph {et~al.}(2012)\citenamefont
  {Tr\"oster}, \citenamefont {Schmitteckert},\ and\ \citenamefont
  {Evers}}]{Troester:PRB:2012}%
  \BibitemOpen
  \bibfield  {author} {\bibinfo {author} {\bibfnamefont {P.}~\bibnamefont
  {Tr\"oster}}, \bibinfo {author} {\bibfnamefont {P.}~\bibnamefont
  {Schmitteckert}}, \ and\ \bibinfo {author} {\bibfnamefont {F.}~\bibnamefont
  {Evers}},\ }\href {\doibase 10.1103/PhysRevB.85.115409} {\bibfield  {journal}
  {\bibinfo  {journal} {Phys. Rev. B}\ }\textbf {\bibinfo {volume} {85}},\
  \bibinfo {pages} {115409} (\bibinfo {year} {2012})}\BibitemShut {NoStop}%
\bibitem [{\citenamefont {Jacob}\ \emph {et~al.}(2009)\citenamefont {Jacob},
  \citenamefont {Haule},\ and\ \citenamefont {Kotliar}}]{Jacob:PRL:2009}%
  \BibitemOpen
  \bibfield  {author} {\bibinfo {author} {\bibfnamefont {D.}~\bibnamefont
  {Jacob}}, \bibinfo {author} {\bibfnamefont {K.}~\bibnamefont {Haule}}, \ and\
  \bibinfo {author} {\bibfnamefont {G.}~\bibnamefont {Kotliar}},\ }\href
  {\doibase 10.1103/PhysRevLett.103.016803} {\bibfield  {journal} {\bibinfo
  {journal} {Phys. Rev. Lett.}\ }\textbf {\bibinfo {volume} {103}},\ \bibinfo
  {pages} {016803} (\bibinfo {year} {2009})}\BibitemShut {NoStop}%
\bibitem [{\citenamefont {Jacob}\ and\ \citenamefont
  {Kotliar}(2010)}]{Jacob:PRB:2010}%
  \BibitemOpen
  \bibfield  {author} {\bibinfo {author} {\bibfnamefont {D.}~\bibnamefont
  {Jacob}}\ and\ \bibinfo {author} {\bibfnamefont {G.}~\bibnamefont
  {Kotliar}},\ }\href@noop {} {\bibfield  {journal} {\bibinfo  {journal} {Phys.
  Rev. B}\ }\textbf {\bibinfo {volume} {82}},\ \bibinfo {pages} {085423}
  (\bibinfo {year} {2010})}\BibitemShut {NoStop}%
\bibitem [{\citenamefont {Jacob}(2015)}]{Jacob:JPCM:2015}%
  \BibitemOpen
  \bibfield  {author} {\bibinfo {author} {\bibfnamefont {D.}~\bibnamefont
  {Jacob}},\ }\href {\doibase 10.1088/0953-8984/27/24/245606} {\bibfield
  {journal} {\bibinfo  {journal} {J. Phys. Condens. Mat.}\ }\textbf {\bibinfo
  {volume} {27}},\ \bibinfo {pages} {245606} (\bibinfo {year}
  {2015})}\BibitemShut {NoStop}%
\bibitem [{\citenamefont {Droghetti}\ and\ \citenamefont
  {Rungger}(2017)}]{Droghetti:PRB:2017}%
  \BibitemOpen
  \bibfield  {author} {\bibinfo {author} {\bibfnamefont {A.}~\bibnamefont
  {Droghetti}}\ and\ \bibinfo {author} {\bibfnamefont {I.}~\bibnamefont
  {Rungger}},\ }\href {\doibase 10.1103/PhysRevB.95.085131} {\bibfield
  {journal} {\bibinfo  {journal} {Phys. Rev. B}\ }\textbf {\bibinfo {volume}
  {95}},\ \bibinfo {pages} {085131} (\bibinfo {year} {2017})}\BibitemShut
  {NoStop}%
\bibitem [{\citenamefont {Stefanucci}\ and\ \citenamefont
  {Kurth}(2015)}]{Stefanucci:NL:2015}%
  \BibitemOpen
  \bibfield  {author} {\bibinfo {author} {\bibfnamefont {G.}~\bibnamefont
  {Stefanucci}}\ and\ \bibinfo {author} {\bibfnamefont {S.}~\bibnamefont
  {Kurth}},\ }\href {\doibase 10.1021/acs.nanolett.5b03294} {\bibfield
  {journal} {\bibinfo  {journal} {Nano Lett.}\ }\textbf {\bibinfo {volume}
  {15}},\ \bibinfo {pages} {8020} (\bibinfo {year} {2015})}\BibitemShut
  {NoStop}%
\bibitem [{\citenamefont {Kurth}\ and\ \citenamefont
  {Stefanucci}(2016)}]{Kurth:PRB:2016}%
  \BibitemOpen
  \bibfield  {author} {\bibinfo {author} {\bibfnamefont {S.}~\bibnamefont
  {Kurth}}\ and\ \bibinfo {author} {\bibfnamefont {G.}~\bibnamefont
  {Stefanucci}},\ }\href {\doibase 10.1103/PhysRevB.94.241103} {\bibfield
  {journal} {\bibinfo  {journal} {Phys. Rev. B}\ }\textbf {\bibinfo {volume}
  {94}},\ \bibinfo {pages} {241103} (\bibinfo {year} {2016})}\BibitemShut
  {NoStop}%
\bibitem [{\citenamefont {Kurth}\ and\ \citenamefont
  {Stefanucci}(2017)}]{Kurth:JPCM:2017}%
  \BibitemOpen
  \bibfield  {author} {\bibinfo {author} {\bibfnamefont {S.}~\bibnamefont
  {Kurth}}\ and\ \bibinfo {author} {\bibfnamefont {G.}~\bibnamefont
  {Stefanucci}},\ }\href {\doibase 10.1088/1361-648X/aa7e36} {\bibfield
  {journal} {\bibinfo  {journal} {J. Phys. Condens. Mat.}\ }\textbf {\bibinfo
  {volume} {29}},\ \bibinfo {pages} {413002} (\bibinfo {year}
  {2017})}\BibitemShut {NoStop}%
\bibitem [{\citenamefont {Jacob}\ and\ \citenamefont
  {Kurth}(2018)}]{Jacob:NL:2018}%
  \BibitemOpen
  \bibfield  {author} {\bibinfo {author} {\bibfnamefont {D.}~\bibnamefont
  {Jacob}}\ and\ \bibinfo {author} {\bibfnamefont {S.}~\bibnamefont {Kurth}},\
  }\href {\doibase 10.1021/acs.nanolett.8b00255} {\bibfield  {journal}
  {\bibinfo  {journal} {Nano Lett.}\ }\textbf {\bibinfo {volume} {18}},\
  \bibinfo {pages} {2086} (\bibinfo {year} {2018})}\BibitemShut {NoStop}%
\bibitem [{\citenamefont {Meir}\ and\ \citenamefont
  {Wingreen}(1992)}]{Meir:PRL:1992}%
  \BibitemOpen
  \bibfield  {author} {\bibinfo {author} {\bibfnamefont {Y.}~\bibnamefont
  {Meir}}\ and\ \bibinfo {author} {\bibfnamefont {N.~S.}\ \bibnamefont
  {Wingreen}},\ }\href {\doibase 10.1103/PhysRevLett.68.2512} {\bibfield
  {journal} {\bibinfo  {journal} {Phys. Rev. Lett}\ }\textbf {\bibinfo {volume}
  {68}},\ \bibinfo {pages} {2512} (\bibinfo {year} {1992})}\BibitemShut
  {NoStop}%
\bibitem [{Note1()}]{Note1}%
  \BibitemOpen
  \bibinfo {note} {This follows by integrating over $\omega $ both sides of
  Eq.~(\ref {main1}), changing variable $\omega '=\omega +V_{{\protect \rm
  T},\protect \rm xc}(\omega )$ in the r.h.s. and taking into account the
  Jacobian $\protect \frac {{\protect \rm d} \omega '}{{\protect \rm d} \omega
  }= 1/ (1 - \protect \frac {\gamma _{{\protect \rm T}}}{\pi } \protect \frac
  {\partial V_{\protect \rm T,xc}(\omega )}{\partial I_T}\unhbox \voidb@x \hbox
  {$\protect \mathcal {A}$}_s(\omega +V_{\protect \rm T,xc}(\omega
  )))$.}\BibitemShut {Stop}%
\bibitem [{\citenamefont {Jakobs}\ \emph {et~al.}(2010)\citenamefont {Jakobs},
  \citenamefont {Pletyukhov},\ and\ \citenamefont
  {Schoeller}}]{JakobsPletyukhovSchoeller:10}%
  \BibitemOpen
  \bibfield  {author} {\bibinfo {author} {\bibfnamefont {S.~G.}\ \bibnamefont
  {Jakobs}}, \bibinfo {author} {\bibfnamefont {M.}~\bibnamefont {Pletyukhov}},
  \ and\ \bibinfo {author} {\bibfnamefont {H.}~\bibnamefont {Schoeller}},\
  }\href {\doibase 10.1103/PhysRevB.81.195109} {\bibfield  {journal} {\bibinfo
  {journal} {Phys. Rev. B}\ }\textbf {\bibinfo {volume} {81}},\ \bibinfo
  {pages} {195109} (\bibinfo {year} {2010})}\BibitemShut {NoStop}%
\bibitem [{\citenamefont {Dittmann}\ \emph {et~al.}(2018)\citenamefont
  {Dittmann}, \citenamefont {Splettstoesser},\ and\ \citenamefont
  {Helbig}}]{dittmann2017non}%
  \BibitemOpen
  \bibfield  {author} {\bibinfo {author} {\bibfnamefont {N.}~\bibnamefont
  {Dittmann}}, \bibinfo {author} {\bibfnamefont {J.}~\bibnamefont
  {Splettstoesser}}, \ and\ \bibinfo {author} {\bibfnamefont {N.}~\bibnamefont
  {Helbig}},\ }\href {\doibase 10.1103/PhysRevLett.120.157701} {\bibfield
  {journal} {\bibinfo  {journal} {Phys. Rev. Lett.}\ }\textbf {\bibinfo
  {volume} {120}},\ \bibinfo {pages} {157701} (\bibinfo {year}
  {2018})}\BibitemShut {NoStop}%
\bibitem [{Note2()}]{Note2}%
  \BibitemOpen
  \bibinfo {note} {Using the Friedel sum-rule one can show that $\unhbox
  \voidb@x \hbox {$\protect \mathcal {G}$}_{\alpha \beta }=C_{\alpha \beta
  }(\gamma _{\protect \rm T},\gamma _{\protect \rm L},\gamma _{\protect \rm
  R})A(\mu )$ where the prefactor $C_{\alpha \beta }$ depends only on the
  hybridizations and $A(\mu )$ is the interacting spectral function at chemical
  potential $\mu $ (which is set to zero in our case). Since $A(\mu )=\protect
  \frac {4}{\gamma }\protect \qopname \relax o{sin}^2(\pi N/2)$ and since in
  i-DFT the KS occupation $N$ is the same as the interacting $N$ we conclude
  that the interacting and KS conductances are the same.}\BibitemShut {Stop}%
\bibitem [{\citenamefont {Motahari}(2017)}]{Motahari:PhD:2017}%
  \BibitemOpen
  \bibfield  {author} {\bibinfo {author} {\bibfnamefont {S.}~\bibnamefont
  {Motahari}},\ }\emph {\bibinfo {title} {Kondo physics and thermodynamics of
  the Anderson impurity model by distributional exact diagonalization}},\
  \href@noop {} {Ph.D. thesis},\ \bibinfo  {school} {Martin-Luther University
  Halle-Wittenberg, Germany} (\bibinfo {year} {2017})\BibitemShut {NoStop}%
\bibitem [{\citenamefont {Requist}()}]{Requist:private:2017}%
  \BibitemOpen
  \bibfield  {author} {\bibinfo {author} {\bibfnamefont {R.}~\bibnamefont
  {Requist}},\ }\href@noop {} {}\bibinfo {note} {{p}rivate
  communication}\BibitemShut {NoStop}%
\bibitem [{\citenamefont {Bertrand}\ \emph {et~al.}(2019)\citenamefont
  {Bertrand}, \citenamefont {Florens}, \citenamefont {Parcollet},\ and\
  \citenamefont {Waintal}}]{Bertrand:arxiv:2019}%
  \BibitemOpen
  \bibfield  {author} {\bibinfo {author} {\bibfnamefont {C.}~\bibnamefont
  {Bertrand}}, \bibinfo {author} {\bibfnamefont {S.}~\bibnamefont {Florens}},
  \bibinfo {author} {\bibfnamefont {O.}~\bibnamefont {Parcollet}}, \ and\
  \bibinfo {author} {\bibfnamefont {X.}~\bibnamefont {Waintal}},\ }\href@noop
  {} {\bibfield  {journal} {\bibinfo  {journal} {arXiv:1903.11646}\ } (\bibinfo
  {year} {2019})}\BibitemShut {NoStop}%
\bibitem [{\citenamefont {Krivenko}\ \emph {et~al.}(2019)\citenamefont
  {Krivenko}, \citenamefont {Kleinhenz}, \citenamefont {Cohen},\ and\
  \citenamefont {Gull}}]{Krivenko:arxiv:2019}%
  \BibitemOpen
  \bibfield  {author} {\bibinfo {author} {\bibfnamefont {I.}~\bibnamefont
  {Krivenko}}, \bibinfo {author} {\bibfnamefont {J.}~\bibnamefont {Kleinhenz}},
  \bibinfo {author} {\bibfnamefont {G.}~\bibnamefont {Cohen}}, \ and\ \bibinfo
  {author} {\bibfnamefont {E.}~\bibnamefont {Gull}},\ }\href@noop {} {\bibfield
   {journal} {\bibinfo  {journal} {arXiv:1904.11527}\ } (\bibinfo {year}
  {2019})}\BibitemShut {NoStop}%
\bibitem [{\citenamefont {Kurth}\ and\ \citenamefont
  {Jacob}(2018)}]{Kurth:EPJB:2018}%
  \BibitemOpen
  \bibfield  {author} {\bibinfo {author} {\bibfnamefont {S.}~\bibnamefont
  {Kurth}}\ and\ \bibinfo {author} {\bibfnamefont {D.}~\bibnamefont {Jacob}},\
  }\href {\doibase https://doi.org/10.1140/epjb/e2018-90184-7} {\bibfield
  {journal} {\bibinfo  {journal} {Eur. Phys. J. B}\ }\textbf {\bibinfo {volume}
  {91}},\ \bibinfo {pages} {101} (\bibinfo {year} {2018})}\BibitemShut
  {NoStop}%
\end{thebibliography}
\end{document}